\documentclass{article}

\usepackage{graphicx} 
\usepackage{subfigure} 

\usepackage{listings}

\usepackage{listings, xcolor}

\definecolor{verylightgray}{rgb}{.97,.97,.97}

\lstdefinelanguage{Solidity}{
	keywords=[1]{anonymous, assembly, assert, balance, break, call, callcode, case, catch, class, constant, continue, contract, debugger, default, delegatecall, delete, do, else, event, export, external, false, finally, for, function, gas, if, implements, import, in, indexed, instanceof, interface, internal, is, length, library, log0, log1, log2, log3, log4, memory, modifier, new, payable, pragma, private, protected, public, pure, push, require, return, returns, revert, selfdestruct, send, storage, struct, suicide, super, switch, then, this, throw, transfer, true, try, typeof, using, value, view, while, with, addmod, ecrecover, keccak256, mulmod, ripemd160, sha256, sha3}, 
	keywordstyle=[1]\color{blue}\bfseries,
	keywords=[2]{address, bool, byte, bytes, bytes1, bytes2, bytes3, bytes4, bytes5, bytes6, bytes7, bytes8, bytes9, bytes10, bytes11, bytes12, bytes13, bytes14, bytes15, bytes16, bytes17, bytes18, bytes19, bytes20, bytes21, bytes22, bytes23, bytes24, bytes25, bytes26, bytes27, bytes28, bytes29, bytes30, bytes31, bytes32, enum, int, int8, int16, int24, int32, int40, int48, int56, int64, int72, int80, int88, int96, int104, int112, int120, int128, int136, int144, int152, int160, int168, int176, int184, int192, int200, int208, int216, int224, int232, int240, int248, int256, mapping, string, uint, uint8, uint16, uint24, uint32, uint40, uint48, uint56, uint64, uint72, uint80, uint88, uint96, uint104, uint112, uint120, uint128, uint136, uint144, uint152, uint160, uint168, uint176, uint184, uint192, uint200, uint208, uint216, uint224, uint232, uint240, uint248, uint256, var, void, ether, finney, szabo, wei, days, hours, minutes, seconds, weeks, years},	
	keywordstyle=[2]\color{teal}\bfseries,
	keywords=[3]{block, blockhash, coinbase, difficulty, gaslimit, number, timestamp, msg, data, gas, sender, sig, value, now, tx, gasprice, origin},	
	keywordstyle=[3]\color{violet}\bfseries,
	identifierstyle=\color{black},
	sensitive=false,
	comment=[l]{//},
	morecomment=[s]{/*}{*/},
	commentstyle=\color{gray}\ttfamily,
	stringstyle=\color{red}\ttfamily,
	morestring=[b]',
	morestring=[b]"
}

\usepackage{fancyhdr}
\usepackage{lipsum}
\usepackage{lastpage}

\usepackage{natbib}

\usepackage{algorithm}
\usepackage{algorithmic}

\usepackage{hyperref}

\usepackage[accepted]{icml2013} 

\icmltitlerunning{Evaluating and Exchanging Machine Learning Models on the Ethereum Blockchain}

\begin{document} 
\twocolumn[
\icmltitle{Trustless Machine Learning Contracts; Evaluating and Exchanging\\
Machine Learning Models on the Ethereum Blockchain}

\icmlauthor{A. Besir Kurtulmus}{besir@algorithmia.com}
\icmladdress{Algorithmia Research,
            1925 Post Alley, Seattle, WA 98101 USA}
\icmlauthor{Kenny Daniel}{kenny@algorithmia.com}
\icmladdress{Algorithmia Research,
            1925 Post Alley, Seattle, WA 98101 USA}

\icmlkeywords{ethereum, blockchain, machine learning, marketplace, trustless evaluation}

\vskip 0.3in
]

\begin{abstract} 
Using blockchain technology, it is possible to create contracts that offer a reward in exchange for a trained machine learning model for a particular data set. This would allow users to train machine learning models for a reward in a trustless manner.

The smart contract will use the blockchain to automatically validate the solution, so there would be no debate about whether the solution was correct or not. Users who submit the solutions won’t have counterparty risk that they won’t get paid for their work. Contracts can be created easily by anyone with a dataset, even programmatically by software agents.

This creates a market where parties who are good at solving machine learning problems can directly monetize their skillset, and where any organization or software agent that has a problem to solve with AI can solicit solutions from all over the world. This will incentivize the creation of better machine learning models, and make AI more accessible to companies and software agents.

A consequence of creating this market is that there will be a well defined price of GPU training for machine learning models. Crypto-currency mining also uses GPUs in many cases. We can envision a world where at any given moment, miners can choose to direct their hardware to work on whichever workload is more profitable: cryptocurrency mining, or machine learning training.
\end{abstract} 

\section{Background}
\label{background}

\subsection{Bitcoin and cryptocurrencies}
Bitcoin was first introduced in 2008 to create a decentralized method of storing and transferring funds from one account to another. It enforced ownership using public key cryptography. Funds are stored in various addresses, and anyone with the private key for an address would be able to transfer funds from this account. To create such a system in a decentralized fashion required innovation on how to achieve consensus between participants, which was solved using a blockchain. This created an ecosystem that enabled fast and trusted transactions between untrusted users.

Bitcoin implemented a scripting language for simple tasks. This language wasn’t designed to be  turing complete. Over time, people wanted to implement more complicated programming tasks on blockchains. Ethereum introduced a turing-complete language to support a wider range of applications. This language was designed to utilize the decentralized nature of the blockchain. Essentially it’s an application layer on top of the ethereum blockchain.

By having a more powerful, turing-complete programming language, it became possible to build new types of applications on top of the ethereum blockchain: from escrow systems, minting new coins, decentralized corporations, and more. The Ethereum whitepaper talks about creating decentralized marketplaces, but focuses on things like identities and reputations to facilitate these transactions. \cite{eth_paper} In this marketplace, specifically for machine learning models, trust is a required feature. This approach is distinctly different than the trustless exchange system proposed in this paper.

\subsection{Breakthrough in machine learning}

In 2012, Alex Krizhevsky, Ilya Sutskever, and Geoff Hinton were able to train a deep neural network for image classification by utilizing GPUs. \cite{imagenet_cnn_paper} Their submission for the Large Scale Visual Recognition Challenge (LSVRC) halved the best error rate at the time. GPU’s being able to do thousands of matrix operations in parallel was the breakthrough needed to train deep neural networks.

With more research, machine learning (ML) systems have been able to surpass humans in many specific problems. These systems are now better at: lip reading \cite{lip_reading_paper}, speech recognition \cite{speach_recog_paper}, location tagging \cite{geo_loc_paper}, playing Go \cite{go_paper}, image classification \cite{human_imagenet_performance_paper}, and more. 

In ML, a variety of models and approaches are used to attack different types of problems. Such an approach is called a Neural Network (NN). Neural Networks are made out of nodes, biases and weighted edges, and can represent virtually any function. \cite{nn_any_func_paper}

\begin{figure}[h]
\caption{Neural Network Schema}
\includegraphics[width=8cm]{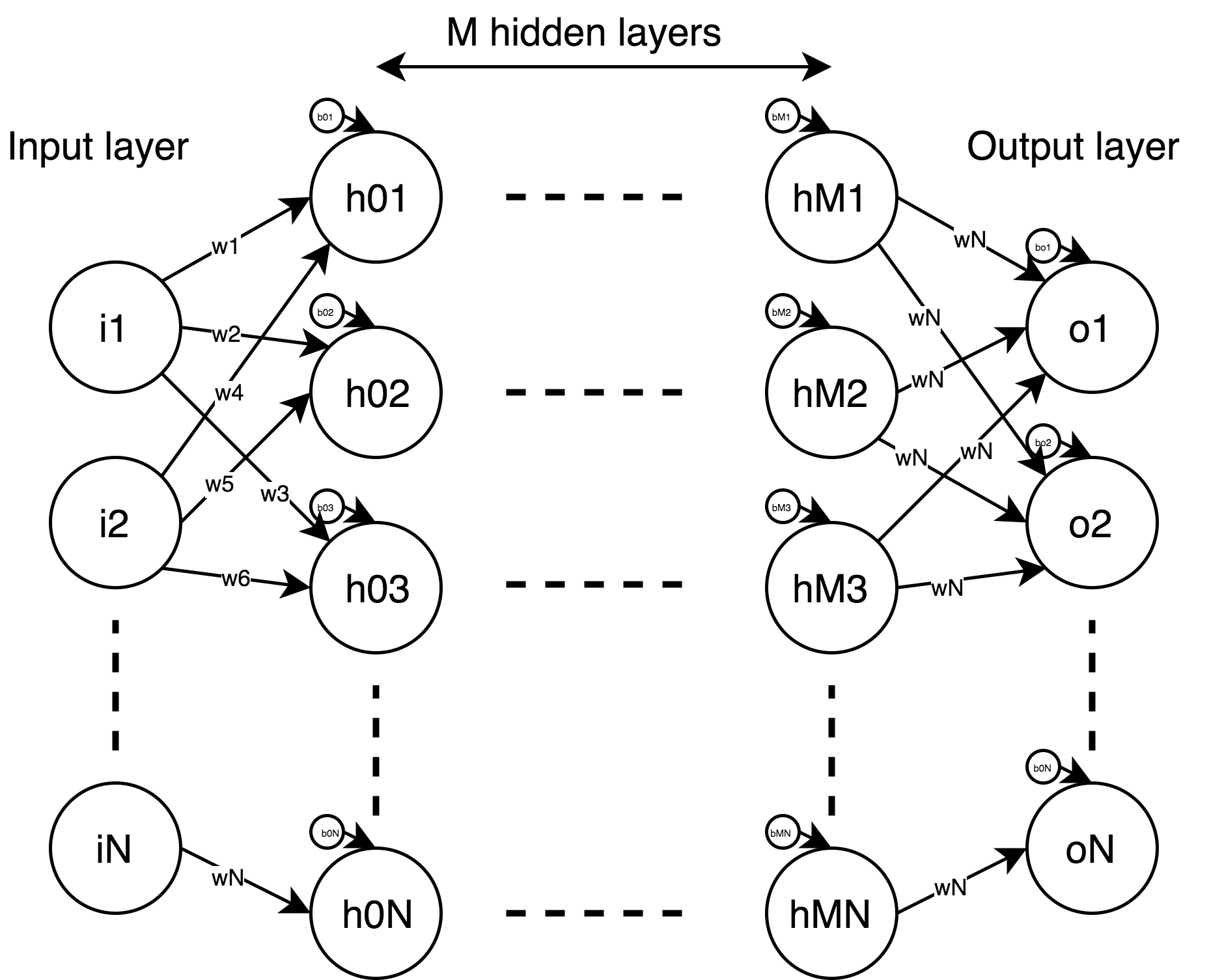}
\end{figure}

There are two steps in building a new machine learning model. The first step is training, which takes in a dataset as an input, and adjusts the model weights to increase accuracy for the model. The second step is testing, that uses an independent dataset for testing the accuracy of the trained model. This second step is necessary to validate the model and to prevent a problem known as overfitting. An overfitted model is very good at a particular dataset, but is bad at generalizing for the given problem.

Once it has been trained, a ML model can be used to perform tasks on new data, such as prediction, classification, and clustering.

There is a huge demand for machine learning models, and companies that can get access to good machine learning models stand to profit through improved efficiency and new capabilities. Since there is strong demand for this kind of technology, and limited supply of talent, it makes sense to create a market for machine learning models. Since machine learning is purely software and training it doesn’t require interacting with any physical systems, using blockchain for coordination between users, and using cryptocurrency for payment is a natural choice.

\section{Introduction}
The ethereum white paper talks about on-chain decentralized marketplaces, using the identity and reputation system as a base. It does not go into detail regarding the implementation, but mentions a marketplace being built on top of concepts like identities and reputations. \cite{eth_paper}

Here we introduce a new protocol on top of the Ethereum blockchain, where identities and reputations are not required to create a marketplace transaction. This new protocol establishes a marketplace for exchanging machine learning models in an automated and anonymous manner for participants.

The training and testing steps are done independently to prevent issues such as overfitting. ML models are verified and evaluated by running them in a forward pass manner on the Ethereum Virtual Machine. Participants using this protocol are not required to trust each other. Trust is unnecessary because the protocol enforces transactions using cryptographic verification.

\section{Protocol}
\label{protocol}

To demonstrate a transaction, a simple Neural Network and forward pass capability is implemented to showcase an example use case.

Basic structure:

\begin{enumerate}
\item \textbf{Phase 1}: User Alice submits a dataset, an evaluation function, and a reward amount to the ethereum contract. The evaluation function takes in a machine learning model, and outputs a score indicating the quality of that model. The reward is a monetary reward, typically denominated in some cryptocurrency (eg- bitcoin or ether).
\item \textbf{Phase 2}: Users download the dataset submitted by user Alice, and work independently to train a machine learning model that can represent that data. When a user Bob succeeds in training a model, he submits his solution to the blockchain.
\item \textbf{Phase 3}: At some future point, the blockchain (possibly initiated by a user action) will evaluate the models submitted by users using the evaluation function, and select a winner of the competition.
\end{enumerate}

\textbf{Note}: Some extra steps are necessary in each of the phases in order to ensure trust and fairness of the competition. Details to follow in section \ref{protocol}.

The DanKu (\textbf{Dan}iel + \textbf{Ku}rtulmus) protocol is proposed to allow users to solicit machine learning models for a reward in a trustless manner. The protocol has 5 stages to ensure a contract is executed successfully.

\begin{figure}[h]
\caption{Initialization stage}
\includegraphics[width=8cm]{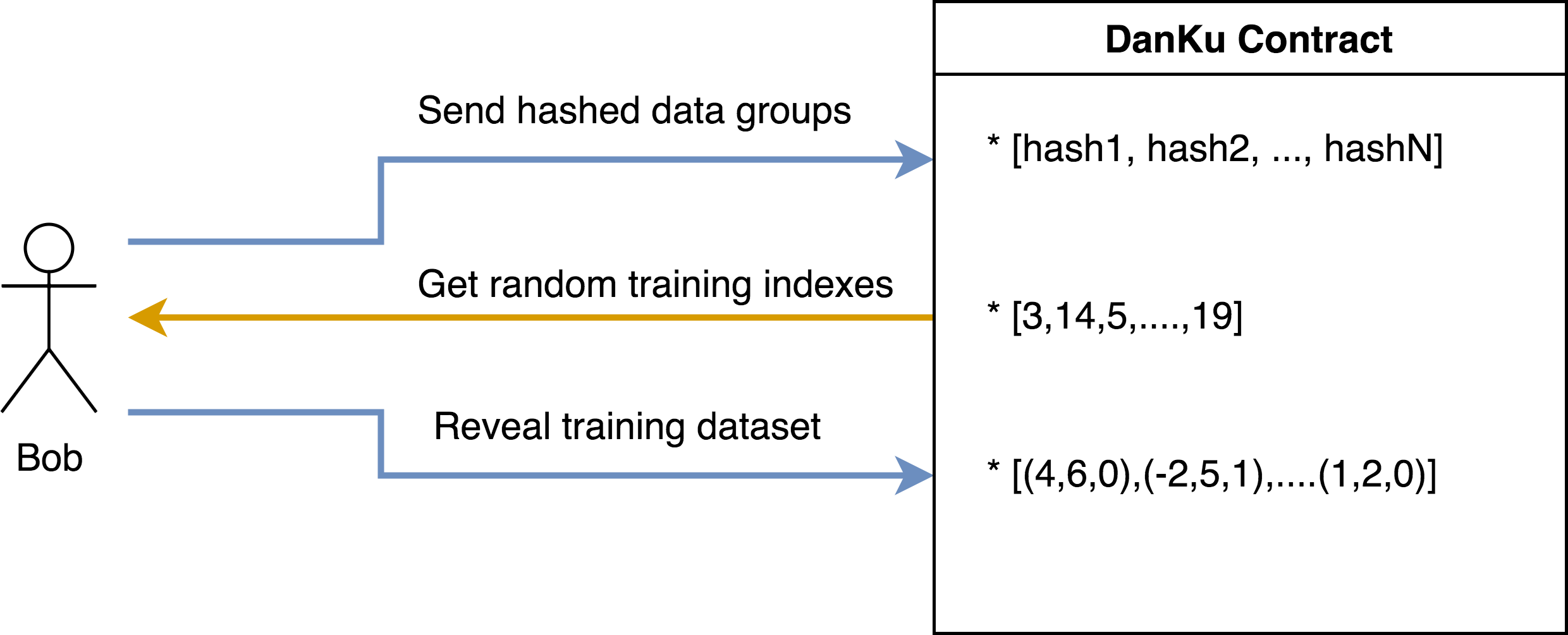}
\end{figure}
\begin{figure}[h]
\caption{Submission stage}
\includegraphics[width=8cm]{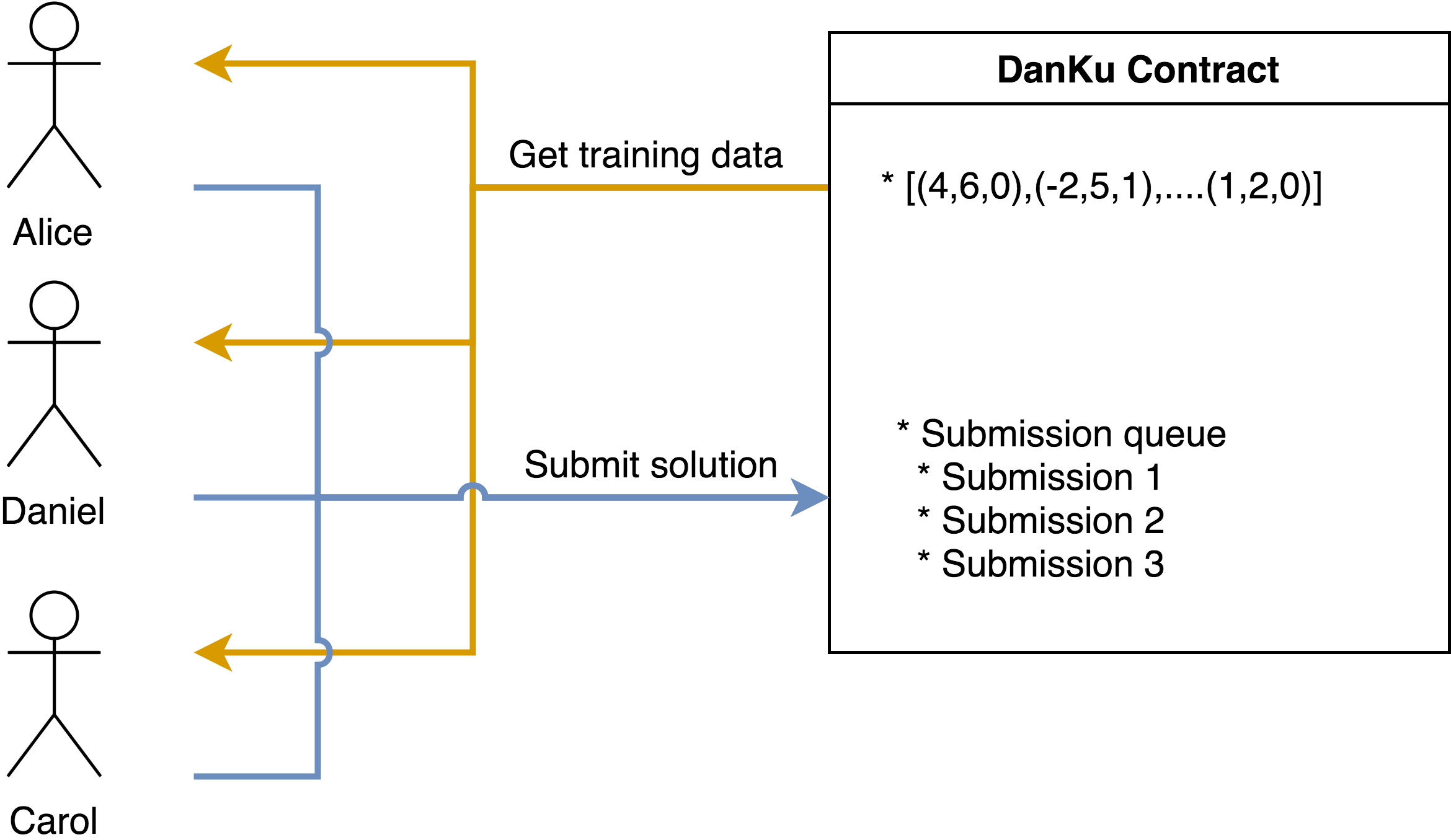}
\end{figure}
\begin{figure}[h]
\caption{Test dataset reveal stage}
\includegraphics[width=8cm]{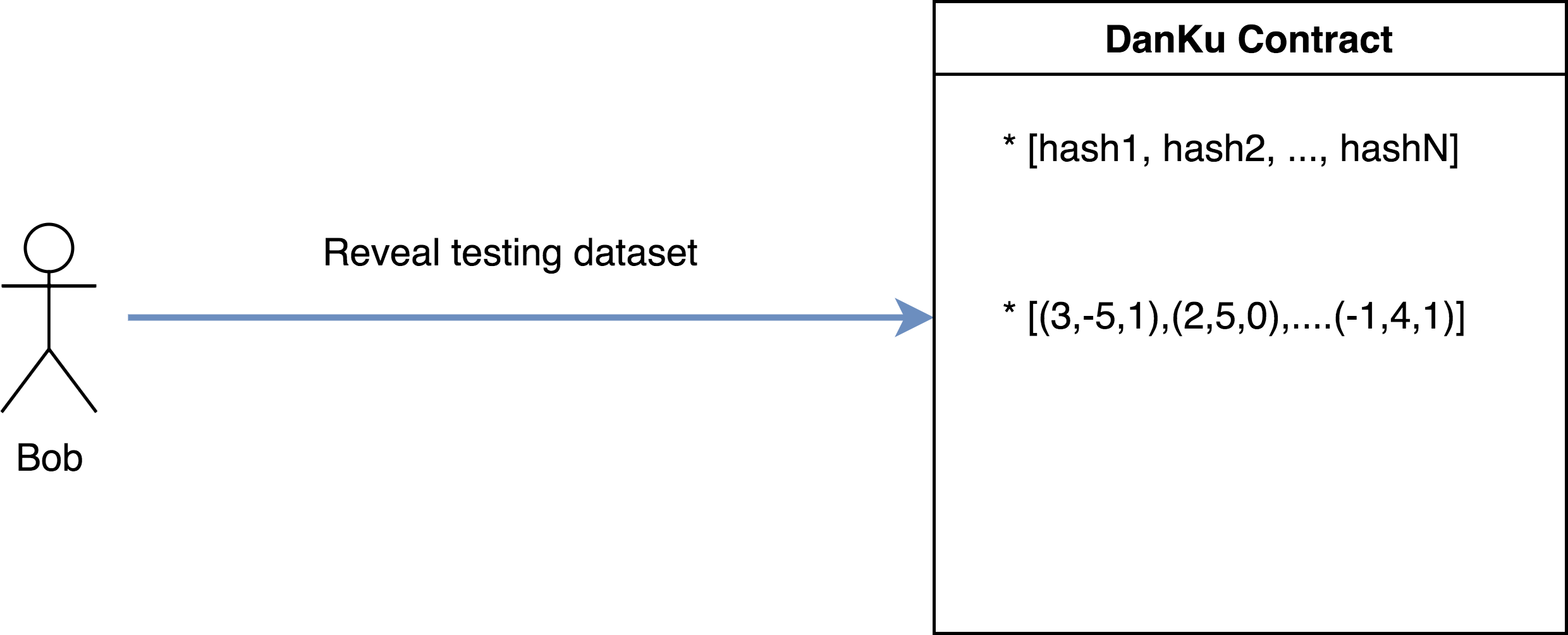}
\end{figure}
\begin{figure}[h]
\caption{Evaluation stage}
\includegraphics[width=8cm]{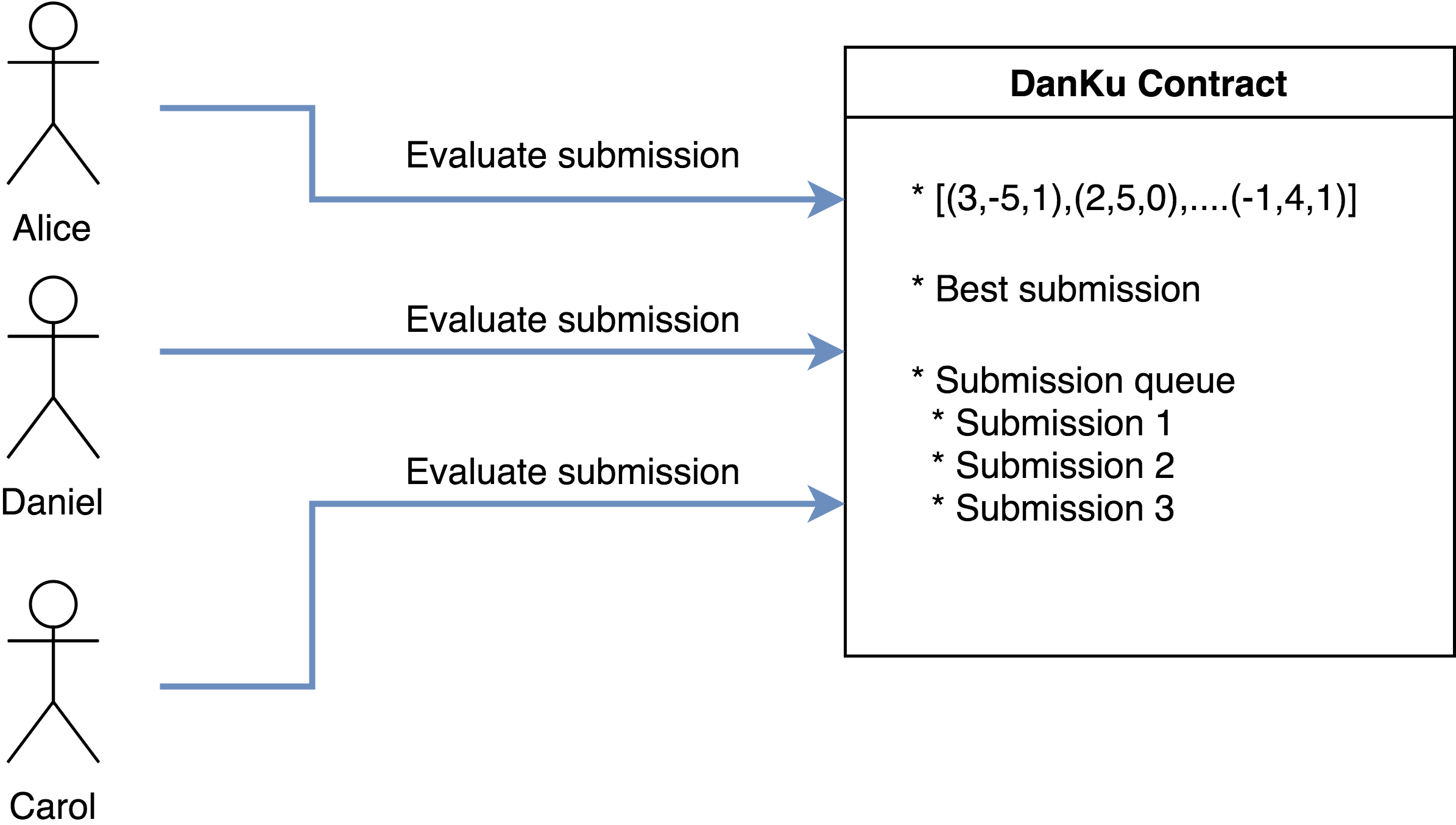}
\end{figure}
\begin{figure}[h]
\caption{Finalize stage}
\includegraphics[width=8cm]{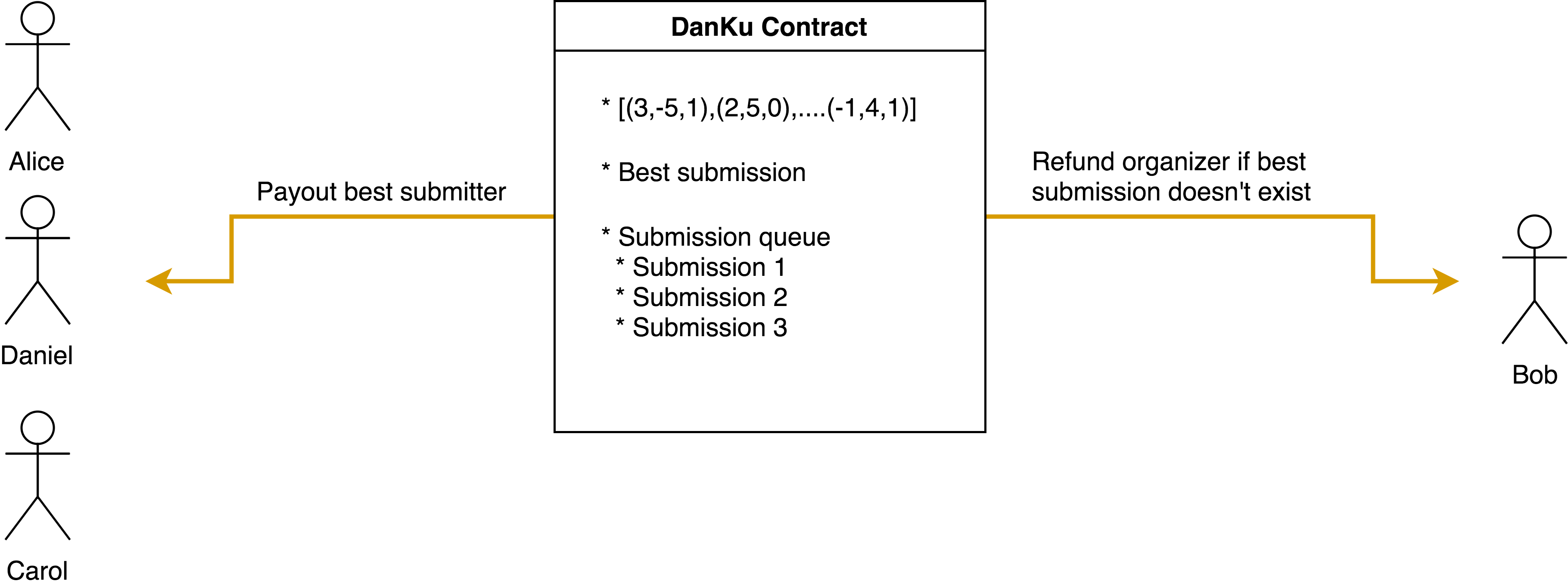}
\end{figure}

\subsection{Definitions}

\begin{enumerate}
    \item A \underline{user} is defined by anyone who can interact with ethereum contracts.
    \item A \underline{DKC (DanKu Contract)} is an Ethereum contract that implements the DanKu protocol.
    \item An \underline{organizer} is defined by a user who creates the DanKu contract.
    \item A \underline{submitter} is a user who submits solutions to the DanKu contract in anticipation for a reward.
    \item A \underline{period} is a timeframe unit that is made up of number of blocks mined.
    \item A \underline{data point} is made up of input(s) and prediction(s).
    \item A \underline{data group} is a matrix made up of data points.
    \item A \underline{hashed data group} is the hash of a data group that also includes a random nonce.
    \item A \underline{contract wallet address} is an escrow account that holds the reward amount until the contract is finalized.
\end{enumerate}

The contract uses the following protocol:

\subsection{Contract Initialization Stage}

An organizer has a problem they wish to solve (expressed as an evaluation function), datasets for testing and training, and an Ethereum wallet.

\begin{enumerate}
\item Contract creation: The organizer creates a contract with the following elements:
    \begin{enumerate}
    \item A machine learning model definition. (a neural network is used in the demo)
    \item \underline{init1()}, \underline{init2()} and \underline{init3()} functions.
    \item Function \underline{get\_training\_index()} to receive training indexes.
    \item Function \underline{get\_testing\_index()} to receive training indexes.
    \item Function \underline{init3()} for revealing the training dataset to all users.
    \item Function \underline{submit\_model()} for submitting solutions to the contract.
    \item Function \underline{get\_submission\_id()} for getting submission id.
    \item Function \underline{reveal\_test\_data()} for revealing the testing dataset to all users.
    \item Function \underline{get\_prediction()} for running a forward pass for a given model.
    \item Function \underline{evaluate\_model()} for evaluating a single model.
    \item Function \underline{cancel\_contract()} for canceling the contract before revealing the training dataset.
    \item Function \underline{finalize\_contract()} for paying out to the best submission, or back to the organizer if best model does not fulfill evaluation criteria...
    \end{enumerate}
\item Init Step 1:
    \begin{enumerate}
    \item Reward deposited in the contract wallet address to payout winning solution.
    \item Maximum lengths of the submission, evaluation and test reveal periods defined by the height of the blocks.
    \item Hashed data groups.
    \end{enumerate}
\item Init Step 2: The organizer triggers the randomization function to generate the indexes for the training and testing data groups. These indexes are selected by using block hash numbers as the seed for randomization.
\item Init Step 3: The organizer reveals the training data groups and nonces by sending them to the contract via the \underline{init3()} function. The data is cryptographically verified by the previously provided hash values for the data groups.
\end{enumerate}

At this point, the DanKu contract is initialized and the training dataset is public. The submitters can download the training dataset to start training their models. After successfully training they can submit their solutions.

\subsection{Solution Submission Stage}

During this phase, any user can submit a potential solution. This is the only period during which submissions will be accepted.

\begin{enumerate}
\item Submitter(s) invoke the \underline{submit\_model()} function, providing a solution with the following elements:
    \begin{enumerate}
    \item The solution weights and biases.
    \item The model definition.
    \item The payment address for payout.
    \end{enumerate}
\end{enumerate}

\subsection{Evaluation Stage}

The evaluation stage can be initiated in one of two ways:

\begin{enumerate}
\item The organizer calls the function \underline{reveal\_test\_data()} that reveals the testing dataset.
    \begin{enumerate}
    \item In this case, the testing dataset will be used for evaluation
    \end{enumerate}
\item The test reveal period has passed, but the organizer has not yet called \underline{reveal\_test\_data()}
    \begin{enumerate}
    \item Since the testing dataset is unavailable, the training dataset will be used for evaluation
    \end{enumerate}
\end{enumerate}

Once the evaluation stage begins, submissions will no longer be accepted, and submitters may now begin evaluating their models:

\begin{enumerate}
\item Submitter(s) call the \underline{evaluate\_model()} function with their submission id.
\item If the model passes the evaluation function, and is better than the best model submitted so far (or is the first model ever evaluated), it is marked as the best model.
\end{enumerate}

\subsection{Post-Evaluation Stage}

\begin{enumerate}
\item After the evaluation period ends, any user can call the \underline{finalize\_contract()} function to payout the reward to the best model submitter.
\item If best model does not exist, reward is paid back to the organizer.
\end{enumerate}

\section{Incentives and Threat Models}
\label{incentives_and_threat_models}

The key to any marketplace is user trust. We need to design a system where no user can cheat or gain advantage over any other user. We must consider this from the point of view of all involved parties: the organizer, the submitters, and even the miners of the ethereum blockchain.

We foresee several risks that need to be mitigated:

\subsection{Overfitting by modeller}

If the submitter has access to the testing dataset, they can overfit their model to this dataset. This is a cardinal sin in machine learning. The submitter can “cheat” by training heavily on the test set, the downside being that the model will not apply to more general problems.

The solution to this problem is to keep the testing dataset secret until the submission period ends. In our contract we require the organizer to come back later and reveal the test set.

\subsection{Test set manipulation by author}
\label{test_set_manipulation_by_author}

If the training and testing set do not draw from the same input distribution, then the organizer can cheat a modeller out of their fee. The organizer can provide fake testing data, and collect the model weights without having to payout to the submitters.

The solution to this problem is to ensure that the organizer cannot pick and choose the training and testing datasets. The organizer submits the hashes of the data groups that represent the whole dataset. The DanKu contract randomly selects a subset of that data that the organizer is obligated to reveal later. Since the hashes are deterministic, the validity of revealed data groups can be verified. This prevents the organizer to manipulate the training/testing dataset.

\subsection{Organizer doesn’t reveal the testing dataset}

For any given reason the organizer may not reveal the testing dataset. This would prevent calling the evaluation function and from paying out submitters for their work in these situations.

To mitigate this issue, we give the organizer a certain amount of time to reveal the testing dataset. If the organizer fails to do so, the submitters can still trigger the evaluation function on the training dataset instead.

Evaluating the models with the training dataset is far from ideal, but it’s a required last resort so submitters would get rewarded for their work in these rare circumstances.

\subsection{Too many submissions}

In the ethereum network all transactions are executed by miners. With every transaction, a fee is required to compensate the miner for the executing and saving it on the blockchain. This fee is called gas. Since each block is mined every 12 seconds on average, it is not practical to accept transactions that would take a long time. Accepting these types of long transactions would increase the risk of missing a block, and not being out-mined by other miners. To mitigate this issue, miners have their own self-imposed gas limits. If a transaction uses more than the gas limit, it will fail regardless of the amount being paid.

Originally the evaluation function was called by the organizer, or submitter if the testing dataset was not revealed. This evaluation function would evaluate all submitted models. If the DanKu contract had too many models to evaluate, the evaluation function was at the risk of running out of gas due to existing gas limits. At the time writing this, the average gas limit was around ~8 million. This would create a vulnerability where the function caller can run out of gas pretty quick, or can get rejected immediately due to the large gas size not getting accepted at any nodes. This would mean that the contract would never be finalized.

To mitigate this problem, we allow users to call the evaluation function on any submitted model. This prevents the too-many-submissions problem since the evaluation function only runs on a single model at a time. This also reduces the amount of total computation needed to evaluate the models. This would create a situation where only the best model submitters would be incentivized to call the evaluation function on their own models.

Since evaluation can also be done offline, submitters can evaluate each model locally to determine if their model is indeed the best one or is in the top-n. It might make sense to call the evaluation function on your model if you’re in the top-n, since there might be a chance that the best model submitter might not call the evaluation function on their model. In an ideal situation this should not happen, but if it does, the next best model submitter can also claim the prize if it ever happens.

\subsection{Rainbow table attacks on hashed data groups}

Initially the sha-3-keccak was only hashing the data group without a nonce. This made this hashing method susceptible to rainbow table attacks from the submitters.

One possible solution to this problem was to implement bcrypt, and use a salted hash instead. Password hashing functions do work, but they’re computationally more expensive. Instead we decided to add a nonce to the end of the data group and hash it using sha3-keccak instead.

This method prevents rainbow table attacks, and keeps the gas cost at a lower price point.

There are probably other solutions to this problem, including like scaling data points and adding noise. This makes sense since real life data is rarely clean, and mostly noisy.

\subsection{Block hash manipulation by miners}

Initially we were only using the last block hash as a source of entropy for our randomization function. The problem with this approach is a miner can influence the resulting hash for a given block. This creates an opportunity for the organizer to cheat as mentioned in section \ref{test_set_manipulation_by_author}. If they’re a miner, they can deploy their contract right after mining a block. This doesn’t guarantee that their contract will be initialized, but it gives them some influence over the selection process for the training and testing dataset.

Being a miner does not give you full control over how the training selection process works. But it would show you which data groups are going to be selected. Due to this, the organizer can decide to not mine a specific block hash that could result in undesirable training indexes.

To minimize the influence of the attacker, we require them to call \underline{init2()} function within 5 blocks of calling \underline{init1()}. If they fail to do so, the contract will get cancelled.

To read more about the implementation of hashing, and the probability of getting favorable training indexes, please refer to section \ref{determining_the_training_dataset}.

\subsection{Distributed reward system abuse}

Depending on a selection criteria, the reward can be claimed by the first submitter who fulfills the evaluation criteria, the best model submitter or both. The reward could even be distributed among top solutions to incentivize more participation. The organizer has full control over picking the selection criteria for their DanKu contract.

A distributed reward system would be similar to Ethereum's proportionate mining reward for stale blocks. A stale block is a solution to a block that is propagated to the network too late. These blocks are still rewarded to make sure miners still get paid for the work they contribute, and to keep the network stable. Only up to 7 stale blocks are rewarded. Stale blocks can happen due to many reasons like outdated mining software, and bad network connectivity.  \cite{eth_paper}

A distributed reward system may de-incentivize submitters since a malicious submitter might resubmit the same solution with minimum changes to solution to steal the work of the original submitter. Due to this reason, it makes sense to only payout to the best submitter. If there are two submissions with the same solution, and this solution is the best one, the first submitter will get paid out instead if they’re both evaluated. This is to prevent malicious submitters from re-submitting the same solution and calling the evaluation function before the original submitter.

The ideal situation would be where these models are trained in pools, and the reward would be distributed to members among a pool when a solution is found. This would make collaboration and distributed rewarding possible. Read more about pool mining in section \ref{gpu_miner_arbitrage}.

\section{Implementation}
\label{inplementation}

\subsection{Hashing the Dataset}

The organizer breaks down the whole dataset into several data groups. A nonce is a randomly generated number that is only intended to be used once. Difference nonces are generated for each data group. Each individual nonce is pushed to the end of the corresponding data group.

The organizer hashes these data groups by passing them to a hashing function. For the DanKu protocol sha3-keccak is chosen as the hashing function for creating hashed data groups.

Ideally each data group is made up of 5 data points. If there is 100 data points, that would make a total of 20 hashed data groups. This later allows the partitioning of the dataset into training and testing datasets.

\subsection{Determining the Training Dataset}
\label{determining_the_training_dataset}

During Initialization step 2, the organizer calls the \underline{init2()} function. This function uses the previously mined blocks hash numbers as a seed for randomization. This function randomly selects the training and testing groups. The default ratio is 80

The randomization function calculates the sha3-keccak of the previous block's hash number, and returns the hexdigest. The modulo of the hexdigest is used to randomly select an index. After selecting an index, we decrement the modulo and repeat the previous step until all training data group indexes are selected. The initial modulo is the total number of data groups.

Algorithm (in solidity) for randomly selecting hashes:

\begin{lstlisting}[language=Solidity]
function randomly_select_index(uint[] array) private {
  uint t_index = 0;
  uint array_length = array.length;
  uint block_i = 0;
  // Randomly select training indexes
  while(t_index < training_partition.length) {
    uint random_index = uint(sha256(block.blockhash(block.number-block_i))) % array_length;
    training_partition[t_index] = array[random_index];
    array[random_index] = array[array_length-1];
    array_length--;
    block_i++;
    t_index++;
  }
  t_index = 0;
  while(t_index < testing_partition.length) {
    testing_partition[t_index] = array[array_length-1];
    array_length--;
    t_index++;
  }
}
\end{lstlisting}

\begin{flushleft}
  Let’s call the number of data groups we have $G$. This makes the probability of getting a favorable index $\frac{1}{G}$.
  This index is selected by getting the modulo $G$ of a given block hash.
\end{flushleft}

\begin{flushleft}
  Let’s call the training percentage as $TP$. The number of training indexes are $G\times TP$.
\end{flushleft}

\begin{flushleft}
  With every selected index, we decrease the modulo $G$ by $1$ for selecting the next index. We iterate to $G\times (1-TP)$ until we have selected all training indexes.
\end{flushleft}

\begin{flushleft}
  Therefore, the probability of getting a unique sequence of training indexes are $\prod\limits_{n=G\times (1-TP) +1}^{G} \frac{1}{n}$
\end{flushleft}

\begin{flushleft}
  A malicious organizer would not be only interested in a sequence of training indexes. This is because any permutation of those indexes would yield the same training dataset. The permutation for the training indexes are: $(G\times TP)!$.
\end{flushleft}

\begin{flushleft}
  This makes the probability of getting ideal training indexes: $(G\times TP)!\times \prod\limits_{n=G\times (1-TP) +1}^{G} \frac{1}{n}$
\end{flushleft}

\begin{flushleft}
  This probability can also be re-written as: $\prod\limits_{n=G\times (1-TP) +1}^{G} \frac{G-n+1}{n}$
\end{flushleft}

\begin{flushleft}
Let’s call the block limit $L$. After calling \underline{init1()}, the organizer has to call \underline{init2()} within $L$ blocks to limit the influence they have over selecting the training indexes. In the DanKu contract, a default of $5$ blocks is used, which should correspond to $1$ minute on average.
\end{flushleft}

Ideally speaking for the organizer, let’s assume that the contract gets deployed immediately. This gives the organizer L chances to call the \underline{init2()} randomization function after \underline{init1()}.

\begin{flushleft}
Hence, this makes the probability of getting ideal training indexes within $L$ blocks: $P = L\times \prod\limits_{n=G\times (1-TP) +1}^{G} \frac{G-n+1}{n}$.
\end{flushleft}

\begin{flushleft}
For a training partition of $80\%$ and a block limit of $5$, here are the chances of getting an ideal group of training indexes for the following number of data groups:
\end{flushleft}

\begin{center}
  \begin{tabular}{ | l | l | }
    \hline
    \# of Data Groups $G$ & Ideal probability $P$ \\ \hline
    5 & $100\%$ \\ \hline
    10 & $~11.11\%$ \\ \hline
    15 & $~1.0989\%$ \\ \hline
    20 & $~0.103199\%$ \\ \hline
    25 & $~0.00941088\%$ \\ \hline
    30 & $~0.00084207\%$ \\
    \hline
  \end{tabular}
\end{center}

As seen in the table, the higher the number of data groups are, the less likely a malicious organizer can affect the outcome.

\subsection{Forward Pass}

A forward pass function is implemented for the given machine learning model definition. For the scope of this paper, we only included a simple neural network definition and a forward pass function. The general idea is to demonstrate that it should be possible to implement some if not most ML functions and models.

\subsection{Evaluating the Dataset}

Depending on the type of problem we’re trying to solve, we can use metrics like accuracy, recall, precision, F1 score, etc. to evaluate the success of the given model.

Based on the scope and requirements, any one of these metrics can be selected. An ideal scoring metric for every classification problem doesn’t exist. Every ML problem should be evaluated within its own scope since it can have sensitivity towards different metrics. (eg. less tolerance towards false-negatives in cancer prediction)

For demonstration purposes, we chose a simple accuracy implementation for evaluating submitted solutions.

\subsection{Math Functions}

The Ethereum Virtual Machine (EVM) doesn’t have a complete math library. It also does not support floating point numbers. The absence of these functions require to implement Machine Learning (ML) models using integer programming, fixed-point float point numbers, and linear activation functions.

Certain activation functions used in ML models such as sigmoid require functions such as \underline{exp()}. These functions also require some floating point constants such as Euler's number e. Both of these math features are not implemented in EVM yet. Implementing them in a contract would significantly increase the gas cost. This makes non-linear type functions less desirable to use in evaluating ML models. For this reason, we’ve used ReLU instead of Sigmoid.

While implementing fixed float point numbers in solidity we noticed that shift functions were implemented with exponentiation functions, therefore they’re actually more expensive than using division.

In solidity the right shift $x>>y$ is equivalent to $\frac{x}{2^y}$. For division, this requires a lot more operations for a simple division. Due to this fact, we’ve extensively used division instead of shifting in fixed float point calculations.

Overall, since the EVM is low level language, and does not have a complete math library, most things needs to be implemented from scratch to get ML models working.

\subsection{Working Around Stack Too Deep Errors}

Solidity only allows to use around 16 local variables in a function. This includes function parameters and the return variable too. Due to this restriction, complex functions such as \underline{forward\_pass()} needed to be divided into several functions.

This required to use minimum number of local variables in functions. Due to this limit, in many places variables were accessed directly instead of be referred by more descriptive variables. This tradeoff was partially mitigated by adding more explanatory comments.

\section{Miscellanea And Concerns}
\label{miscellanea_and_concerns}

\subsection{Complete anonymization of the Dataset and Model Weights}

Currently neither the dataset or model weights and biases are anonymized. Any user can access the submitted models on DanKu contracts.

A possible way to solve this issue might be through the use of homomorphic encryption. \cite{ML_encrypted_data_paper} One thing that stands out with homomorphic encryption is the use of real numbers. Since integers are real numbers, this would work with our implementation in solidity. This implementation also uses integers to compute the forward pass.
With homomorphic encryption, it’s possible to encrypt the data set for the sake of privacy. \cite{homomorphic_encryption_paper} It’s also possible to encrypt the weights of a model that is trained on an un-encrypted dataset. \cite{safe_ai_article} One thing to keep in mind is that even though homomorphic encryption can provide anonymity, it’ll make the contract more expensive to execute. For the scope of this paper, homomorphic encryption is not included in the protocol.

\subsection{Storage gas costs and alternatives}

It is known that storing datasets in the contract and validating them would require significant amounts of gas.

Here’s a solidity contract that writes 1 KB of data to the ethereum blockchain on solidity version 0.4.19:

\begin{lstlisting}[language=Solidity]
pragma solidity ^0.4.19;

contract StorageTest {
    byte[1024] data;
    function store() public {
        for (uint i = 0; i < 1024; i++) {
            data[i] = 'A';
        }   
    }
}
\end{lstlisting}

After creating the contract, the transaction cost for storing 1 KB of data is about 6068352 gas. This is currently below the gas limit of 8 million (as of Jan 2018). This means that it is possible to write large datasets in increments instead of a single transaction.

MNIST is a popular handwritten digits dataset used for benchmarking optical character recognition algorithms. The whole dataset is around 11594722 bytes. \cite{mnist_dataset_url} The average gas price is around 4 gwei (as of Jan 2018). This makes the total cost of writing the MNIST dataset around 275 ethereum. As of writing, ethereum is worth around \$1,100 (Jan 2018). This would make the total cost about \$302,500.

The gas price is mostly consistent across different solidity versions for the same code instructions. The total price can change over time, since it’s determined by the gas and ethereum price.

A big portion of ML problems definitely have larger datasets than MNIST. Storing these datasets in the blockchain isn’t a sustainable solution in majority of cases.

Alternatives like IPFS and swarm might be used for storing the datasets elsewhere. These alternatives try to tackle the problem of storing large files and datasets on the blockchain, while keeping the price at a reasonable level.

\subsection{Model execution timeout}

Like with all other ethereum contracts, there’s always a gas limit for running DanKu contracts. Some complex models may not run due to high gas costs. Miners might reject these models if it requires gas more than the limit imposed by the miner. Accepting to execute such a model would make mining a block more likely to become stale.

Even though the gas limit goes up on average, the limit itself prevents running a set of deep neural networks. This means that running certain models may not be possible until the gas limit goes up, or EVM gets further optimized.

\subsection{Re-implementation of Offline Machine Learning Libraries}

Since solidity only works with integers, most popular machine learning libraries won’t work out-of-the-box with these contracts. These libraries might need to be adapted to work with integers instead of floating points.

This might be possible if the activation functions are linear and the weights and biases are integers.

\subsection{Cancelling a contract}

Organizers are allowed to cancel a contract and take back their reward if they haven’t revealed their training dataset yet.

\section{Other Considerations and Ideas}
\label{other_considerations_and_ideas}

\subsection{GPU miner arbitrage}
\label{gpu_miner_arbitrage}

A consequence of creating this market is that there will be a well defined price of GPU training for machine learning models. Crypto-currency mining also uses GPUs in many cases. We can envision a world where at any given moment, miners can choose to direct their hardware to work on whichever workload is more profitable: cryptocurrency mining, or machine learning training.

GPU miners who choose to join these mining pools will be joining these pools that will be managed by Data Scientists. If that pool solves the contract, ideally the reward will be divided between the Data Scientists who manage the pool and the miners who provide the hardware.

Additionally the way to verify proof-of-work is a lot more simpler than in traditional cryptocurrency mining. In a pool a submitted solution and its accuracy can be considered the unit of work, which would make it easier to assess the amount of work done by each individual miner. Each unit of work can be easily assessed via the evaluation function.

\subsection{Self-improving AI systems}

Furthermore, we envision a world where intelligent systems can use these contracts to improve their own capabilities, by requesting training on new problem domains.

Since these contracts use cryptocurrencies to reward participants, all interactions with these contracts are digital, hence ideal for self-improving AI systems.

\subsection{Raising Money for Computation Power for Medical Research}

DanKu contracts can also be used for medical research purposes. This has the advantage of being able to directly donate to the contract wallet address. This removes the requirement for a middleman or trusting a 3rd party. As the reward gets bigger, it’ll attract more participants to submit solutions to the given problem.

For example, a DanKu contract could be created for a protein-folding problem, that might help with cancer research. This would create a new way to crowdsource funds for medical research.

\subsection{Potential improvements}

It’s worth mentioning that the DanKu protocol has a lot of room for improvement. As mentioned before, introducing homomorphic encryption would definitely be a useful feature. Better designed DanKu contracts could significantly reduce gas costs. \cite{solidity_optimization_paper} The solidity language could introduce new features that would make DanKu contracts faster and cheaper. The ever increasing gas limit will make running certain machine learning models possible that weren’t before. Improvements in Machine Learning such as using 8-bit integers will help further reduce gas costs for DanKu contracts. \cite{tpu_paper}

And possibly, a new language specifically designed for matrix multiplication for the ethereum blockchain can significantly increase performance for DanKu contracts.

\section{Conclusion}

The DanKu protocol is a byproduct of two emerging technologies that are disrupting their respective fields. It utilizes the anonymous and distributed nature of smart contracts, and the intelligent problem solving aspect of machine learning. It also introduces a new method for crowdsourcing funds for computational research.

The protocol helps users solicit machine learning models for a given fee. The protocol does not require trust and works completely on a decentralized blockchain.

The protocol creates interesting opportunities like GPU mining arbitrage, provides a more transparent platform for raising money for things like medical research, and introduces an automated self-improvement system for AI agents.

It is expected that open-source ML models will significantly benefits from this. We might see a sudden rise of publicly available ML models available in the open-source community.

The protocol will potentially create a new marketplace where no middlemen is required. It’ll further democratize machine learning models, and increase opportunity in acquiring these models. This new levelled playing field should hopefully benefits both blockchain technology and machine learning, as it’ll provide a more efficient mean of obtaining machine learning models and increase smart contract usage over time.

\bibliography{example_paper}
\bibliographystyle{icml2013}

\end{document}